\documentclass[fleqn,usenatbib]{mnras}

\usepackage{newtxtext,newtxmath}

\usepackage[T1]{fontenc}
\usepackage{ae,aecompl}

\usepackage{graphicx}	
\usepackage{amsmath}	
\usepackage{amssymb}	

\usepackage{soul}

	\newcommand{\cdthirty}{\mbox{\,CD\,$-30^{\circ}11223$}}
    \newcommand{\jsixteen}{\mbox{\,J162256+473051}}

\title[Tidal Interactions of Close Hot Subdwarf Binaries]{Tidal Interactions in Post Common-Envelope sdB Binaries}

\author[H. P. Preece, C. A. Tout \& C. S. Jeffery]{
Holly P. Preece,$^{1,2}$\thanks{E-mail: }
Christopher A. Tout$^{1}$ and 
C. Simon Jeffery$^{2}$
\\
$^{1}$Institute of Astronomy, University of Cambridge\\
$^{2}$Armagh Observatory and Planetarium\\
}

\date{Accepted XXX. Received YYY; in original form ZZZ}

\pubyear{2018}

\begin{document}
\label{firstpage}
\pagerange{\pageref{firstpage}--\pageref{lastpage}}
\maketitle

\begin{abstract}
Over half of all observed hot subdwarf B (sdB) stars are found in binaries, and over half of these  are found in close configurations with orbital periods of 10$ \,\rm{d}$ or less. 
In order to estimate  the companion masses in these predominantly single-lined systems, tidal locking has frequently been assumed for sdB binaries with periods less than half a day. 
Observed non-synchronicity of a number of close sdB binaries challenges that assumption and hence provides an ideal testbed for tidal theory. 
We solve the second-order differential equations for detailed 1D stellar models of sdB stars to obtain the tidal dissipation strength and hence to estimate the tidal synchronization time-scale owing to Zahn's dynamical tide. 
The results indicate synchronization time-scales longer than the sdB lifetime in all observed cases. 
Further, we examine the roles of convective overshooting and convective dissipation in the core of sdB stars and find no theoretical framework in which tidally-induced synchronization should occur.  
\end{abstract}




\section{Introduction}
Hot subdwarf B (sdB) stars are compact sub-luminous stars. They have surface temperatures between $20\,000$ and $40\,000\,\rm{K}$ and surface gravities $5<\log_{10}(g_{\rm{surf}}/\rm{cm\,s^{-2})}<6$. The sdBs were first observed by \cite{discoveryzwicky} and their spectra were quantified by \cite{specdefinition}. The stars are helium core burning with low-mass hydrogen envelopes. Typically the stars spend around $150$ $\rm{M \, yrs}$ in their He burning phase. They are thought to be the cores of red giant branch (RGB) stars exposed by close binary-star interaction \citep{han1}. One of the proposed mechanisms for sdB formation is common envelope ejection. The sdBs produced in this manner are in binary systems with orbital periods less than $10\,\rm{d}$. Observations suggest that about half of the observed sdB systems lie in such configurations  \citep{napiow04binaryf,binaryfraction2011}. 

The close sdB binaries are spectroscopically single-lined with either white dwarf (WD) or low-mass main-sequence dM companions. Eclipsing post-common envelope sdBs with a dM companion are referred to as HW Vir type systems. Unless it is eclipsing it is generally not possible to find the inclination of a system. This means it can be difficult to estimate the component masses. By assuming tidal synchronization, a spectroscopic measurement of the projected rotation velocity and an assumed radius constrains the rotational period, the orbital inclination, and hence the companion mass. \cite{methodtsyncfirst} first applied this method to the subdwarf O star HD49798. \cite{geier2010} further applied the same technique to a sample of 51 close sdB stars.

The fact that so many sdBs are in close binaries makes them an ideal test bed for tidal dissipation theories. These theories have always been controversial for stars with convective cores and radiative envelopes such as sdBs. Two competing theoretical prescriptions for dissipation in such stars are given by \cite{zahn75,zahn77} and \cite{tassoul}.  \cite{zahnbad} demonstrated that the dynamical tide proposed by \cite{zahn77} is too inefficient to describe the observed level of synchronization of some early main-sequence spectroscopic binaries, particularly when the fractional radius of the convective region is below 0.05. \cite{tassoul} address this efficiency issue by suggesting that pumping across the Ekman boundary provides a mechanism for tidal dissipation. \cite{tassoulbad} dispute the physical validity of Tassoul's mechanism.

In Section 2 we review the current observations and previous calculations of tidal synchronization. In Section 3 we address the methods used for this paper, first by reviewing tidal theory, then by discussing the numerical methods and finally the stellar models. In Section 4 we present our results and Section 5 concludes.

\section{Observations and Previous Studies of Tidal Synchronization Time-scales}
In a circular orbit, full tidal synchronization has been achieved when the entire star rotates as a solid body with a spin period equal to the binary orbital period. Until recently, only the surface rotation of sdBs could be determined from rotational broadening of spectral lines. The metal lines are used to determine the projected equatorial speed $v_{\rm{rot}}\sin i$. Pulsations of sdBs have been both predicted \citep{sdbpulsepredict} and observed \citep{sdbpulseobs} making them candidates for asteroseismology. If the stars are rotating, one of the degeneracies of the pulsations is broken. This manifests itself as a small symmetric splitting of the pulsation modes. If this splitting can be resolved the internal rotation rate of the star can be determined. If the rotation period and orbital period are known tidal synchronization can be confirmed or dismissed. 

\subsection{Observational Context}

\label{sec:obs} 
\begin{figure}
	\includegraphics[width=\columnwidth]{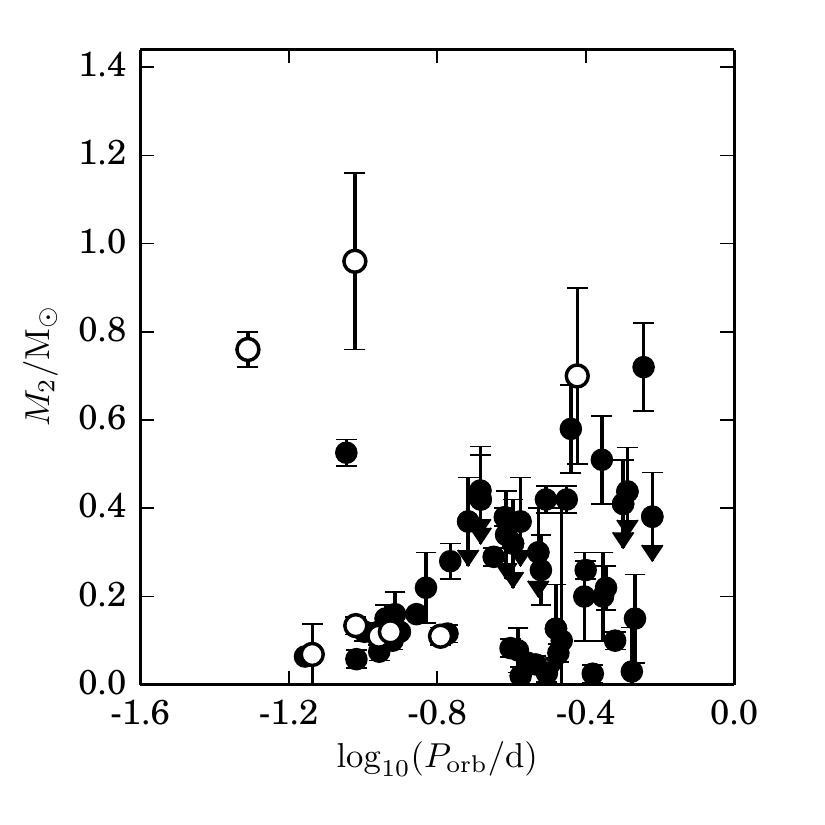}
    \caption{Companion mass, $M_2$, versus orbital period for all observed sdB binary systems with orbital period less than $0.6\,\rm{d}$. The masses of these systems have been estimated from spectroscopic orbits. The white points have assumed tidal synchronization.}
    \label{fig:obs}
\end{figure}
\noindent About 65 sdB binaries with orbital periods below $0.6\,\rm{d}$ have been observed so far. The orbital periods and companion masses of this sample are shown in Fig. \ref{fig:obs}. Of this sample some have assumed tidal synchronization. The observations are summarised by \cite{kupferobs} and in papers cited therein. Several pulsating sdBs have been observed with the \textit{Kepler} mission \citep{keplerpulsators}. The observed properties of the sdBs most relevant to this study are summarised in Table 1. 

\begin{table}
\caption{Rotation and orbit properties of sdBs with orbital periods below $0.6\,\rm{d}$ and known spin periods.}
\begin{tabular}{ |c|c|c| }
\hline
Name & $P_{\rm{orb}}$/$\rm{d}$ & $P_{\rm{rot}}$/$\rm{d}$\\
\hline
\cdthirty & 0.0489 & 0.0427 or 0.0646\\
\jsixteen & 0.069789 & 0.1151563523\\
NY Vir & 0.101016 & 0.101016\\
Feige 48 & 0.34375 & ?\\
KIC 11179657 & 0.394454167 & 7.4\\
B4 & 0.3985 & 9.63 \\
KIC 02991403 & 0.443075 & 10.3\\
\hline
\end{tabular}
\end{table}

Of all observed sdB binaries,  NY Vir is the only object for which the outer layers show evidence of  synchronous rotation with the binary orbit from asteroseismology \citep{charpinetnyvir}. If the star is in synchronous rotation it is a fast rotator which complicates asteroseismological analysis. \cite{charpinetnyvir} obtained an sdB mass of $0.459\pm 0.006\,\rm{M_\odot}$ from asteroseismology. \cite {nyvirsyncmaja} solved for the binary properties of the system using multi-band photometric lightcurves and radial velocity curves from high-resolution spectra. Owing to the correlation between the large number of free parameters and degeneracies in the mass ratio of the binary, three equally probable solutions were obtained. These three solutions predict sdB masses of $0.530$, $0.466$ or $0.389\,\rm{M_\odot}$. The companion mass $M_2$ is either $0.11$ or $0.12\,\rm{M_\odot}$ and the orbital period of the binary $P_{\rm{orb}} = 0.101016\, \rm{d}$. \cite{nyvirmass13}'s seismic analysis measured an sdB mass of $0.471\,\pm \, 0.006\,\rm{M_\odot}$.

Feige 48 was initially thought to be synchronized with $P_{\rm{rot}}\,=9.02\,\pm \, 0.07\,\rm{hr}$ \citep{feige48syncupdated} and $P_{\rm{orb}}\,=9.0\,\pm \, 0.5\,\rm{hr}$ \citep{otoolefeige48}. This $P_{\rm{rot}}$ was determined from asteroseismology with a 6 night campaign at CFHT. \cite{feige48nonsync} remeasured $P_{\rm{orb}}\,=8.24662 \, \rm{hr}$ which challenges conclusion of tidal synchronization. In addition, \cite{fontainefeige48} carried out an extensive 5-month asteroseismic campaign which challenges the $P_{\rm{rot}}$ obtained by \cite{feige48syncupdated}. The true $P_{\rm{rot}}$ remains unknown.

Rotational splitting was measured for the three HW Vir type systems B4 \citep{pablob4}, KIC 02991403 and KIC 11179657 \citep{pablo2nonsync}. All of these were found to be rotating substantially sub-synchronously. B4 is a sdB binary in the NGC 6791 open cluster. It has $P_{\rm{orb}}\,=0.3985\,\rm{d}$ and $P_{\rm{rot}}\,=9.63\,\rm{d}$. The companion has been identified as a low-mass main-sequence star but its mass has not been further constrained. $\rm{KIC}\, 11179657$ has $P_{\rm{orb}}\,=\,9.4669\,\rm{hr}$, $P_{\rm{rot}}\,=\,7.4\,\rm{d}$ and $M_2\,<\,0.26\,\rm{M_\odot}$. $\rm{KIC}\, 02991403$ has $P_{\rm{orb}}\,=\,10.6338\,\rm{hr}$, $P_{\rm{rot}}\,=\,10.3\,\rm{d}$ and $M_2\,<\,0.26\,\rm{M_\odot}$. 

The remaining  sdB binaries observed with {\it Kepler} and with asteroseismically inferred rotation rates are PG1142-037 \citep{pg1142-037}, KIC$\,$7664467 \citep{kic7664467}, KIC$\,$10553698 \citep{kic10553698}, KIC$\,$7668647 \citep{kic7668647}. These have  $13\, \rm{hr}<P_{\rm{orb}}<14\, \rm{d}$ and   $35\, \rm{d} < P_{\rm{rot}} <47\, \rm{d}$. Typical rotation rates for sdBs, without considering the effects of common envelope evolution, have been approximated with measurements from red clump stars, which are considered to have a similar evolutionary origin \citep{redclumprot}. If we take initial spin periods from those of the red clump stars as lying between 30 and 300$\,\rm{d}$, the binaries in wider orbits aren't spun up while those in systems with $P_{\rm{rot}}\ll 30\,\rm{d}$ are somewhat spun up but predominantly not synchronized. 

Further insight is provided by \jsixteen, the shortest period HW\,Vir system known, with $P_{\rm{orb}}\,=0.069789\, \rm{d}$ \citep{j1622+4730}. The system is eclipsing, so the inclination is known and the surface rotation rate can be directly measured from the line profiles. Combined with the measured radius, $P_{\rm{rot}}=0.1151563523\, \rm{d}$ and so \jsixteen\ is rotating non-synchronously. The mass of the sdB star was found to be between 0.28 and 0.64$\,\rm{M_\odot}$, with  $M_{\rm sdB} = 0.48\, \pm\,0.03\,\rm{M_\odot}$ giving the best results \citep{j1622+4730}. With the orbit fully solved, the mass of the unseen companion is found to be $0.064\,\rm{M_\odot}$, well below the H-burning threshold. This is therefore evidence that sub-stellar companions can provide enough energy to remove the H-envelope during common envelope evolution but not enough torque to synchronize the sdB star.

To date, the shortest period sdB binary is \cdthirty\ with $P_{\rm{orb}}\,=0.0489\,\rm{d}$. This system is eclipsing and displays clear signs of ellipsoidal variations. Spectroscopically, the projected surface rotation $v_{\rm{rot}}\sin i= 177\pm 10\,\rm{km\,s^{-1}}$ and the inclination $i = 83.8^{\circ} \pm 0.6$ \citep{cd3011223}. The logarithmic surface gravity $\log (g_{\rm{surf}}/\rm{cm\,s^{-2}})$ of this sdB has been measured as 5.72 from high dispersion spectra and 5.36 from low dispersion spectra \citep{cd-30gsurf1}. The higher solution gravity is consistent with the system being synchronized but the lower is not. If the canonical mass of $0.47\,\rm{M_\odot}$ is assumed for the sdB star, the companion mass $M_2\, = 0.74\,\rm{M_\odot}$. A sdB mass of $0.54 \,\rm{M_\odot}$ and companion mass of $M_2\, = 0.79\,\rm{M_\odot}$ also provide a consistent solution. Because \cdthirty\ is an extreme system with both a short orbital period and a high companion mass, it is the sdB binary most likely to have been synchronized. 

\subsection{Previous Calculations}
\label{sec:previouscalc}
\begin{figure}
	\includegraphics[width=\columnwidth]{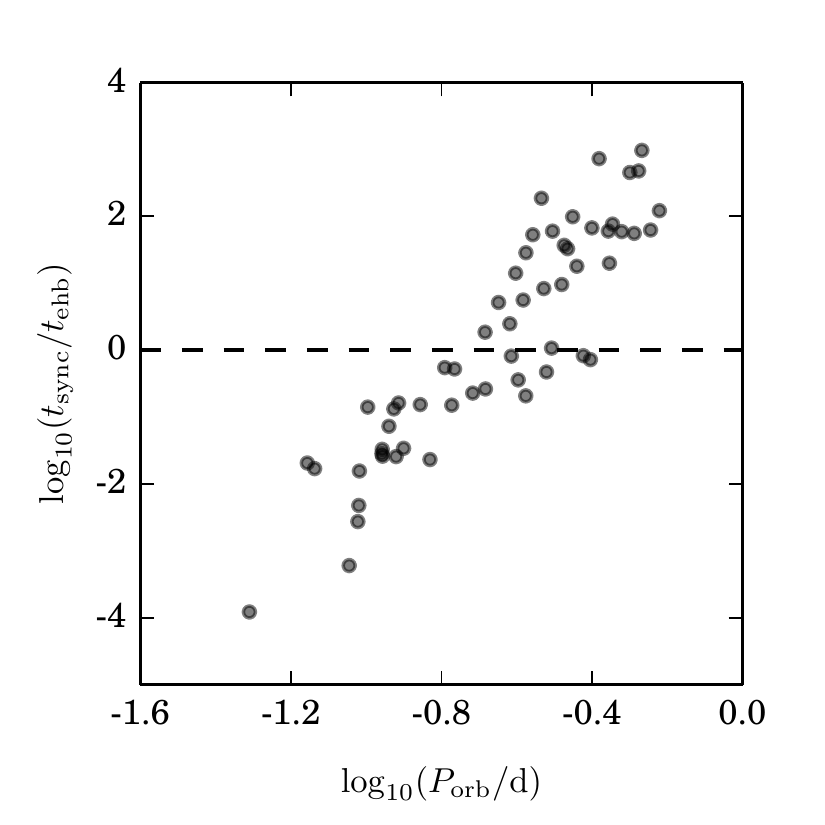}
    \caption{The ratio of synchronization time-scale to the extreme horizontal branch lifetime as a function of the orbital period for the known close sdB binaries with orbital periods less than $0.6\,\rm{d}$ as calculated by Geier et al. (2010) using Zahn's mechanism. Geier's calculations of Zahn's dynamical tide suggest that sdB stars synchronize with orbital periods less than $0.39\,\rm{d}$.}
    \label{fig:geier2010}
\end{figure}

\cite{geier2010} investigated whether the assumption of tidal synchronization could be used to determine the inclination and thus yield the companion mass for close spectroscopically single lined sdB binaries. They analysed a sample of 51 observed sdB stars in binaries with periods below $10\,\rm{d}$. They calculated synchronization time-scales with the theoretical prescriptions described by \cite{zahn77} and \cite{tassoul}. Fig. \ref{fig:geier2010} applies the calculations of synchronization due to Zahn's dynamical tide to the set of known sdBs with orbital periods less than $0.6\, \rm{d}$ assuming an EHB lifetime of 150 $\rm{Myr}$.

The Tassouls' mechanism for dissipation predicts that all systems with $P_{\rm{orb}}<10\,\rm{d}$ are synchronized which is not observed. Zahn's theory of dynamical tides describes tidal dissipation for stars with convective cores and radiative envelopes. The synchronization time-scales depend on the tidal coupling coefficient $E_2$ which is highly dependent on the structure of the star. The coefficient is laborious to calculate so \cite{geier2010} used a scaling from main-sequence models and $E_2$ was approximated as $(r_{\rm{conv}}/R_{\rm{sdB}})^8$ \citep{claretcunha97}, where $r_{\rm{conv}}$ is the radius of the convective core and $R_{\rm{sdB}}$ is the total radius of the sdB star. Note that $r_{\rm{conv}}$ includes any semi-convective region. The sdB model used had $r_{\rm{conv}}/R_{\rm{sdB}}\,=\,0.15$ and a canonical mass of $0.47\,\rm{M_\odot}$. The radii of the sdBs are calculated from observed spectroscopic $g_{\rm{surf}}$. Zahn's dynamical tide doesn't consider dissipation via turbulent convection, this must be calculated separately. See Sec. 3.2 and Sec. 3.3 for details of these mechanisms. 

\cite{geier2010}'s calculations of Zahn's dynamical tide predicted that systems with orbital periods less than $0.39\,\rm{d}$ would synchronize within the EHB lifetime. The study found that the systems with orbital periods up to $1.2\,\rm{d}$ could be solved consistently under the assumption of tidal synchronization. However, using this approach they found a dearth of systems at high inclinations and also predicted some very large companion masses. The assumption of tidal locking is further contradicted by \cite{j1622+4730} and the three Pablo observations. Follow up observations of some of the \cite{geier2010} systems has shown that the observed companion masses are lower than those predicted.

In light of the asteroseismological results for sub-synchronously rotating sdB systems with orbital periods substantially below $1.2\,\rm{d}$, \cite{pablothesis} recalculated the time-scales predicted by Zahn's dynamical tide. This approach was to solve the two required structural differential equations to get a precise $E_2$. He did this for one detailed stellar model with a mass of $0.478\,\rm{M_\odot}$, radius $0.298\,\rm{R_\odot}$ and $r_{\rm{conv}}/R_{\rm{sdB}}\,=\,0.08$ and an undisclosed radius of gyration. These calculations found $E_2$ to be significantly smaller than $(r_{\rm{conv}}/R_{\rm{sdB}})^8$. Ultimately he predicted that systems with $P_{\rm{orb}}<3.6\,\rm{hr}$ should be synchronized within a typical sdB lifetime of $150\,\rm{Myr}$. Pablo's study  looked at only one sdB model which has a fairly large radius compared to most sdBs and only considered dissipation owing to excited, and subsequently damped, g-modes.

\section{Methods}
Calculation of the tidal effects for all dissipation mechanisms considered in this paper requires solving structural differential equations for detailed stellar models. A grid of stellar models was created for this purpose and differential equation solvers were written and included in the tidal dissipation calculation code.

\subsection{Theory}
\begin{figure}
    \includegraphics[width=\columnwidth]{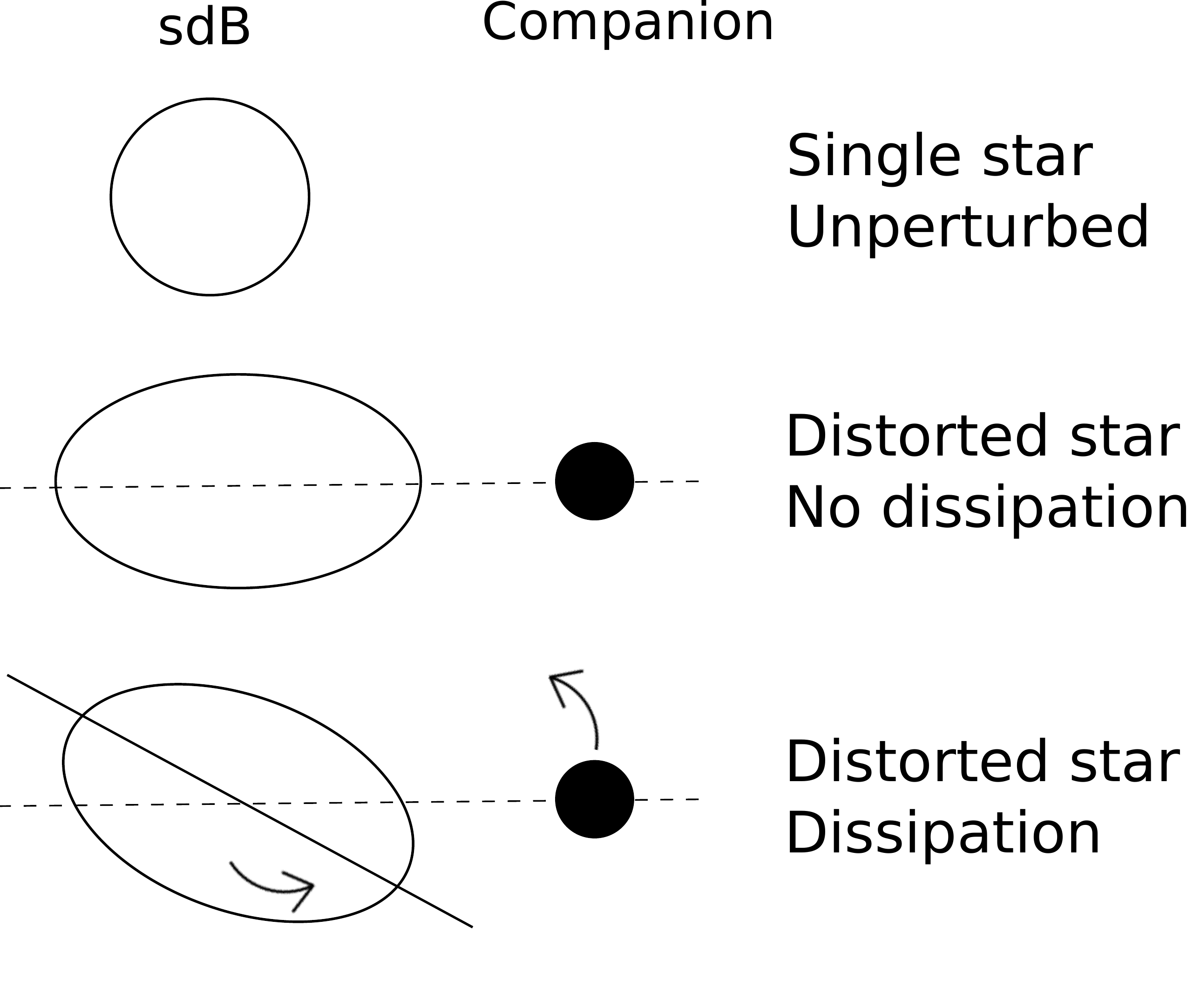}
    \caption{Schematic diagram illustrating basic tidal interactions. The top panel is a single unperturbed star. In the second panel, the star has a close companion causing tidal distortion in the form of a bulge on both sides of the star. This system is either locked or has no dissipation because the bulge is along the line connecting the centres of mass of the objects. The bottom panel shows a tidally distorted star with some sort of dissipation mechanism causing the bulge to move away from the line connecting the centres of mass of the two stars. This system is rotating sub-synchronously, spinning slower than it orbits, causing the tidal bulge to lag behind the line connecting the two centres of masses. The tidal bulges then experience a torque that serves to drive the system to synchronism.}
    \label{fig:tidesdiagram}
\end{figure}

The basic idea of tidal interactions (Fig. \ref{fig:tidesdiagram}) is as follows. If a companion is close to the star there is a difference in the potential between the side closest to and that furthest from the companion star. This causes a bulge to form along the line connecting the centre of masses of the two stars. The star is distended both towards the companion, because the matter there is pulled towards the companion, and away from the companion, because the matter there is less tightly bound.

If the system is synchronized the bulge stays in the same place on the star and always points towards the companion. If the system is not synchronized and there is no dissipation mechanism, the bulge moves around the star always pointing towards the companion. If the system is not synchronized and there is a dissipative mechanism, the bulge moves away from the line connecting the centre of masses. This creates a torque through the star causing it to spin up or down until it is synchronized, if such a stable configuration exists.  

\subsection{Convective Dissipation}
In convective regions the bulk movement of material over large distances causes a natural turbulent viscosity. This viscosity provides a drag which prevents the bulge moving instantaneously around the star and instead introduces a lag. 

Closely following the formalism of \cite{eggstheory} and \cite{eggzbook}, the tidal bulge is modelled as symmetric with a shape described by a second order Legendre polynomial. The mean radius $\bar{r}$ is constant over an equipotential of a tidally distorted star and is defined so that
\begin{equation}
\bar{r}\,=\,r(1+\alpha(r)P_2(\cos\theta)),
\end{equation}
where $r(\theta)$ is radius, $P_2$ is the second order Legendre polynomial and $\theta$ is the polar angle subtended in the star. The amplitude of the distortion at a given radius of the star is defined by  $\alpha$ which is dimensionless and highly dependent on the structure of the star. It can be found by solving Clairaut's second order differential equation,
\begin{equation}
    \alpha'' + \bigg(\frac{4}{r} - \frac{2\phi''}{\phi'}\bigg)\alpha' + \bigg(\frac{2\phi''}{r\phi'}-\frac{2}{r^2}\bigg)\alpha = 0,
\end{equation}
where $\phi$ is the gravitational potential of the star and primes denote derivatives with respect to the radius of an undistorted model $r=\bar{r}$. In the homogeneous density case $\alpha(r)\,=\,\rm{const}$. In the non-homogeneous case Clairaut's equation can be solved for an unperturbed, non-rotating, detailed 1D stellar model, such as is output by the STARS code (see section 3.5). The solution for $\alpha$ then describes a 2D distorted star. The shape of which can be modelled with a quadrupole tensor $q_{ij}$. The moment of this quadrupole tensor is denoted $Q$ and defined by 
\begin{equation}
    Q\,= \,\int_0^R\frac{4\pi\rho r^4(5\alpha + r\alpha')}{5MR^2\alpha(R)} dr,
\end{equation}
where $\rho(r)$ is the density of the star at a given $r$, $\alpha(R)$ is $\alpha(r)$ at the surface of the star and $R$ is the total radius of the star. 

If the orbit of the binary is not synchronized with the rotation, a time varying velocity field is produced within the star as tides are raised and lowered. A model of the interior of the star as a fluid with constant density along equipotentials and a tidal bulge described by a Legendre polynomial allows the velocity field to be described with the equation of continuity. The shape of the tidal distortion can be re-written as 
\begin{equation}
\bar{r}\,=\,r+\,\frac{\alpha(r)}{r}H
\end{equation}
where $H(r,\theta)$ is a harmonic function describing the shape of the distorted star and $\bar{r}$ is constant in $\theta$,
\begin{equation}
H(r,\theta)\,=\,\frac{r^2P_2(\cos\theta)}{d^3},
\end{equation}
with $\pmb{d}$ being the centre of mass of the system and $d\,=\,|\pmb{d}|$. With the time derivative of $H$ denoted by $K$, the continuity equation can be satisfied by by the velocity field $\pmb{v}$ defined as  
\begin{equation}
    \mathbf{v}\,=\,-\frac{1}{2}\beta (r) \alpha (R) \nabla K. 
\end{equation}
Here
\begin{equation}
    \beta (r)\,=\, -\frac{1}{\rho}\int^R_r \frac{\alpha(r)}{\alpha(R)}\frac{d\rho}{dr}dr.
\end{equation}
From mixing length theory, the local turbulent viscosity can be approximated as $\nu = wl$ where $w$ is the mean velocity of the turbulent eddies and $l$ is the mixing length which refers to the size of the largest cells. As the tidal bulge moves around the star, this viscosity provides a dissipative mechanism for the tides. The rate of dissipation of the mechanical energy $\epsilon$ through the star is 
\begin{equation}
    -\frac{d\epsilon}{dt}\,=\,\frac{1}{2}\int\rho w l t_{ij}^2 dV\,=\,\frac{9M_2^2R^6}{2M_1^2(1-Q)^2}s_{ij}^2\int_0^{M_1}wl\gamma(r)dm.
\end{equation} 
The rate of strain tensor is described by $t_{ij}$ and $s_{ij}$ is the symmetric, time dependent, space independent stress tensor. The masses of the primary and secondary stars are given by $M_1$ and $M_2$ respectively and 
\begin{equation}
    \gamma (r) \,=\,\beta^2+\frac{2}{3}r\beta\beta ' +\frac{7}{30}r^2\beta'^2.
\end{equation}
The viscous time-scale of the convective region $\tau_{\rm{visc}}$ is defined by 
\begin{equation}
    \frac{1}{\tau_{\rm{visc}}}\,=\,\frac{1}{M_1R^2_1}\int_0^{M_{1}}wl\gamma(r)dm.
\end{equation}
Care must be taken here to evaluate this only in the convective regions of the star. The tidal time-scale can be found to be
\begin{equation}
   \tau_{\rm{tide}}\,=\,\frac{2\tau_{\rm{visc}}}{9}\frac{a^8}{R^8}\frac{M_1^2(1-Q)^2}{M_2(M_1+M_2)}.
\end{equation}
From this tidal time-scale, the rate of change of rotational angular velocity $\frac{d\Omega}{dt}$ can be found to be
\begin{equation}
    \frac{d\Omega}{dt}\, = \,\frac{\omega}{\tau_{\rm{tide}}}\bigg(1-\frac{\Omega}{\omega}\bigg)\frac{M_2}{M_1+M_2}\frac{a^2}{R^2k_r^2}.
\end{equation}
The radius of gyration of the star $k_{\rm{r}}^2$ refers to the distribution of the components of an object around its rotational axis. It is defined so that $k_{\rm{r}}^2\,=\,I/MR^2$ where $I$ is the moment of inertia of the star. Solving this first order differential equation allows $\Omega (t)$ to be found. From this, the time taken to arrive at a synchronous state can be calculated as
\begin{equation}
    \tau_{\rm{sync}}\,=\,\log\bigg(\frac{\omega-\Omega_0}{\omega-\Omega}\bigg) \frac{\tau_{\rm{tide}}(M_1+M_2)R^2k_r^2}{M_2a^2}.
\end{equation}
It is assumed that the orbital angular velocity $\omega$ remains constant over these time-scales because the moment of inertia of an sdB star is small compared to that of the binary orbit. 
\subsection{Zahn's Mechanism of Radiative Dissipation}
\cite{zahn75,zahn77} developed a theory of dynamical dissipation for stars with radiative envelopes and convective cores. The periodic tidal potential induced by the companion star resonates with g-modes in the core. At the radiative boundary, these excited g-modes are damped. This provides a mechanism for tidal dissipation. The resultant characteristic synchronization time-scale is given by 
\begin{equation}
    \frac{1}{\tau_{\rm{sync}}}\,=\,5\cdot 2^{5/3} \bigg(\frac{g_{\rm{surf}}}{R}\bigg)^{1/2} k_r^2 \bigg(\frac{R}{a}\bigg)^{17/2}q^2(q+1)^{5/6}E_2,
\end{equation}
where $g_{\rm{surf}}$ is the surface gravity of the star and the mass ratio of the stars is  $q\,=\,M_2/M_1$. The tidal coefficient $E_2$ describes the coupling between the tidal potential and the excited pulsations. It is highly dependent on the structure of the star and is defined as
\begin{equation}
    E_2\,=\,\frac{3^{8/3}\Gamma(\frac{4}{3})^2}{(2n+1)(2(2+1))^{4/3}}\frac{\rho R^3}{M}\bigg(\bigg( \frac{N\,^2}{x\,^2}\bigg)^{'}_{\rm{cc}}\frac{\rho R^3}{ \,g_s}\bigg)^{-1/3}H_2^2,
\end{equation}
where $\Gamma(\frac{4}{3})\,=\,0.48060041894$ and $x$ is the fractional radius $r/R$. The Brunt-Vaisala frequency $N^2$ characterises the buoyancy of material within the star. The primes denote derivatives with respect to $x$. The subscript $\rm{cc}$ refers to the convective boundary location. The quantity $H_2$ is 
\begin{equation}
    H_2\,=\frac{2\times 2+1}{((n-3)Y(1)+Y\,'(1))X(x_{\rm{cc}})}\int_0^{x_{\rm{cc}}}\bigg(Y\,''-2(2+1)\frac{Y}{x^2}\bigg)Xdx.
\end{equation}
where $Y$ is the solution to the differential equation
\begin{equation}
    Y\,''-\frac{6}{x}\bigg(1-\frac{\rho}{\bar{\rho}}\bigg)Y\,'-\bigg(6-12(1-\frac{\rho}{\bar{\rho}})\bigg)\frac{Y}{x^2}\,=\,0,
\end{equation}
which is evaluated throughout the star, and $X$ is a structural quantity given by the solution to the differential equation
\begin{equation}
    X\,''-\frac{\rho'}{\rho}X\,'-\frac{6}{x^2}X\,=\,0,
\end{equation}
which is only evaluated in the convective region. This description of tidal dissipation doesn't consider the effect of the convective dissipation. This theory again models the tidal bulge as a second order Legendre polynomial. 

\subsection{Differential Equation Solvers}
Both dissipation prescriptions require solutions to second-order differential equations for detailed 1D stellar models. Both differential equations are initial value problems that can be solved with integrator methods. We constructed an Euler solver, second order Runge-Kutta solver and a fourth order Runge-Kutta solver based on the algorithms presented by \cite{nummethods}. These methods all allow for variable step sizes so errors introduced by interpolation can be avoided. The Euler solution is the fastest computationally but also the least accurate. However all the methods predicted the same time-scales to within $0.5\,$per cent.

\subsection{Stellar Models}
All the stellar models used were created with the Cambridge STARS code \citep{eggz71}. STARS has been modified substantially since its inception \citep{stancliffe09}. It uses OPAL II type opacity tables, allows for binary evolution and follows the chemical evolution of $^1$H, $^3$He, $^4$He, $^{12}$C, $^{14}$N, $^{16}$O and $^{20}$Ne. The code uses an adaptive non-Lagrangian mesh. Convection is treated with mixing-length theory (MLT) as described by \cite{mlt} and uses a MLT parameter of $\alpha\,=\,2.0$ (defined as the ratio of mixing length to pressure scale-height). Semi-convection is treated as a diffusive process \citep{starssemiconv}. Convective overshooting as described by \cite{starsovershoot} is also included. Mass loss on the red giant branch (RGB) is described by Reimers' prescription \citep{reimers}. The sdB star models were made with the method described by \cite{hu10} as follows. 

\subsubsection{The He Flash}
\begin{figure}
	\includegraphics[width=\columnwidth]{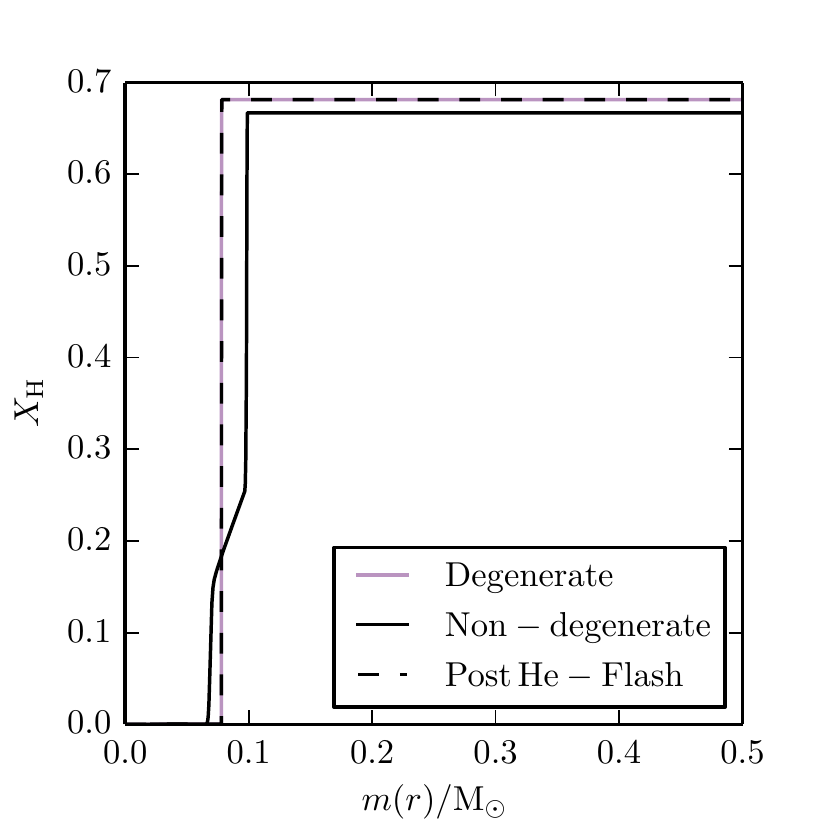}
    \caption{The H composition profile through the post He-flash model. The purple line is the profile of the degenerate star just before He ignition. The black line is the profile of the He burning non-degenerate star before modifying the core or envelope. The purple points show the adjusted 1.75$\,\rm{M_\odot}$ post He-flash model. The adjusted profile maintains the steep composition gradient formed during the red giant branch phase.}
    \label{compadjust}
\end{figure}

The STARS code cannot evolve stars through the He-flash independently. To imitate this process, a star with just enough mass to ignite He quietly and non-degenerately is created \citep{heflashstars}. This is allowed to evolve until just after He is ignited. Next, mass is removed from the star to give it the desired mass. The composition profile through the star is modified and the core is allowed to grow a little to give the same envelope profile and core mass as its degenerate counterpart. The H composition profile for the 1.75$\,\rm{M_\odot}$ post He-flash star is displayed in Fig. \ref{compadjust}. This method works on the principle that once He is ignited, the degeneracy of the core is lifted, and also that the stellar structure is independent of the evolutionary history.
\subsubsection{The sdB Stars}
\begin{figure}
	\includegraphics[width=\columnwidth]{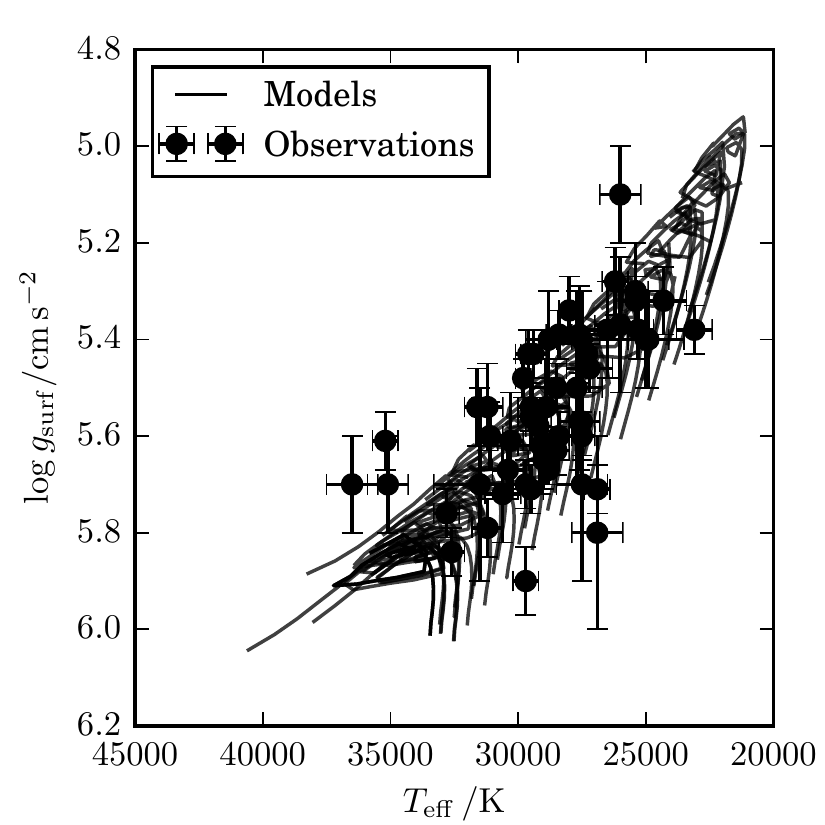}
    \caption{The logarithm of the surface gravity as a function of effective temperature $T_{\rm{eff}}$. The tracks are the sdB models. The observed quantities for the known close sdB binaries shown in Fig. 1 are plotted in black with error bars. The sdBs with the largest envelope masses have the lowest effective temperatures and surface gravities.}
    \label{fig:tefflogg}
\end{figure}

\begin{figure}
	\includegraphics[width=\columnwidth]{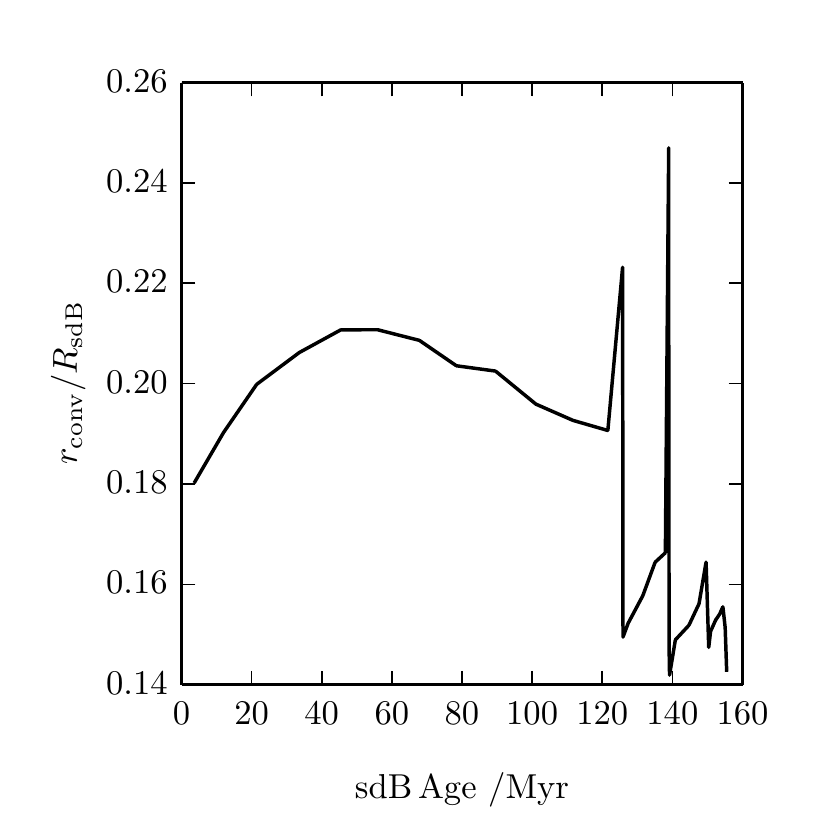}
    \caption{The evolution of the fractional radial extent of the convective core over the sdB evolution for a single evolutionary sequence against time $t$ measured from the start of the sdB phase. Towards the end of the sdB evolution core breathing pulses can be seen. The model is a 0.47$\,\rm{M_\odot}$ sdB star with a $10^{-4}\,\rm{M_\odot}$ envelope.}
    \label{fig:rconvoverr}
\end{figure}

The sdBs were made from three different mass progenitors, $1.25,\, 1.5$ and $1.75\,\rm{M_\odot}$. These stars had respective core masses of 0.4680, 0.4614 and 0.4510$\,\rm{M_\odot}$ at the tip of the RGB. Common envelope ejection was simulated with high mass-loss rates. During the common-envelope simulation the nuclear reactions were turned off and the star was kept in thermal equilibrium. The mass loss was stopped with envelope masses distributed between $0$, where the hydrogen mass fraction reached 0.1, and $0.02\,\rm{M_\odot}$ giving a range of resultant sdB masses distributed near the canonical mass of $0.47\,\rm{M_\odot}$. Each sdB model was then allowed to relax on to the zero-age extreme horizontal branch (ZAEHB) and then to evolve through He burning. As discussed by \cite{schindler}, overshooting affects the mass of the sdB's convective region during its evolution. Models with no overshooting and an overshooting parameter of $\delta_{\rm{ov}}\,=0.12$ \citep{dovstars} were created. This results in a grid of over 800 stellar models with a range of envelope masses and evolutionary states. Fig. \ref{fig:tefflogg} shows the sdB models on a $T_{\rm{eff}}-\log{g}$ diagram with the observed close sdBs from Fig. \ref{fig:obs}. The radial growth of the convective core for a single sdB evolutionary sequence can be seen in Fig. \ref{fig:rconvoverr} for an sdB model with $M_{\rm{sdB}}\,=\, 0.47\,\rm{M_\odot}$ and envelope mass $10^{-4}\, \rm{M_\odot}$ and no convective overshoot.

\section{Results}
We calculated synchronization time-scales for all of the modelled sdBs with companion masses below the Chandrasekhar mass limit and orbital periods less than $4\,\rm{hr}$ for dissipation by Zahn's prescription. Our results suggest that the sdBs cannot become tidally synchronized within the extreme horizontal branch (EHB) lifetime. Traditionally the dynamical tide assumes no dissipation via the equilibrium tide. In the case of sdBs this assumption may not be valid. The sdBs have had the majority of their envelope removed meaning that the convective core now occupies a much more substantial fraction of the star. The synchronization time and change in the rotational period due to the equilibrium tide is also calculated.

\subsection{Zahn's Dynamical Tide}
\begin{figure}
	\includegraphics[width=\columnwidth]{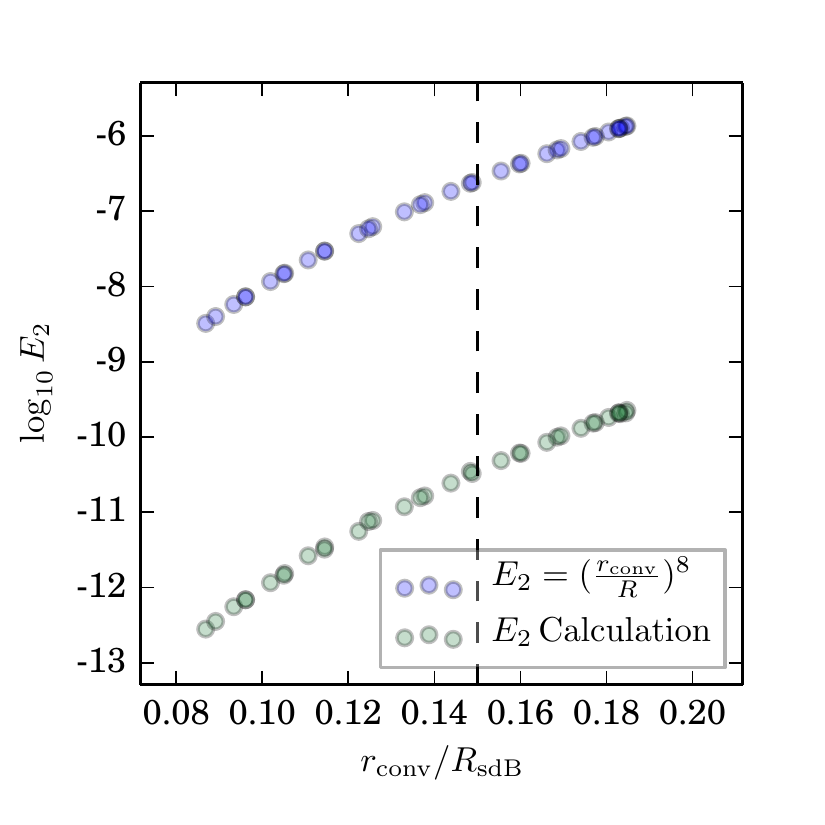}
    \caption{$E_2$ as a function of the ratio of the radial extent of the convective zone of the ZAEHB sdB models. The blue circles are $E_2$ calculated as $(r_{\rm{conv}}/R_{\rm{sdB}})^8$ and the green circles are $E_2$ calculated by solving the required second order differential equations. The blue line is that used for Geier (2010)'s study. The two different methods for calculating $E_2$ give results that differ by an average of 4 orders of magnitude and show that scaling from main-sequence models doesn't work.}
    \label{fig:e2}
\end{figure}

Previous studies of tidal synchronization for sdB stars have focused on Zahn's prescription of tidal dissipation which applies to stars with convective cores and radiative envelopes. The grid of models discussed in section 3.5 was used to solve Eqs. 17 and 18 and then to find the tidal coefficient $E_2$. The results of these calculations are shown in Fig. \ref{fig:e2}. This re-calculation of $E_2$ shows that the main-sequence scaling treatment is a poor approximation. The parametrization over-predicts $E_2$ by at least a factor of 3000. In addition, $E_2$ is highly sensitive to the relative size of the convective region and spreads over two orders of magnitude.

\begin{figure}
	\includegraphics[width=\columnwidth]{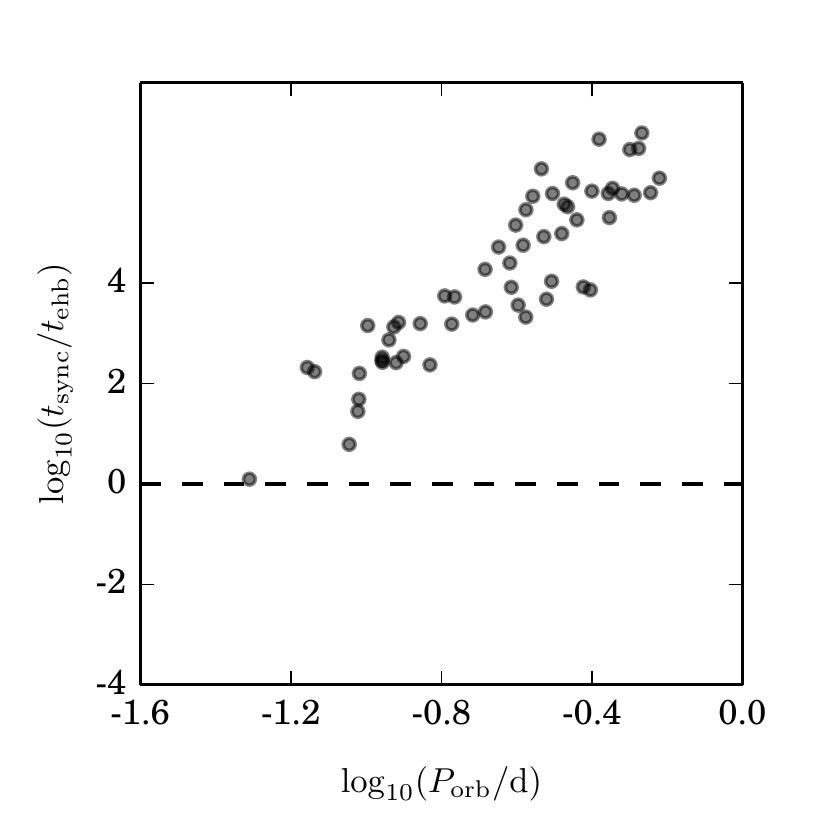}
    \caption{As Fig. 2 with $t_{\rm{sync}}$ computed via Zahn's mechanism with $E_2\,=\,10^{-10.6}$. }
    \label{geier2010mye2}
\end{figure}

\cite{geier2010} assumed $r_{\rm{conv}}/R_{\rm{sdB}}\,=\,0.15$. The ZAEHB sdB model with $r_{\rm{conv}}/R_{\rm{sdB}}\,=\,0.15$ has  $E_2 = 10^{-10.6}$. The results of applying this to the synchronization calculations can be seen in Fig. \ref{geier2010mye2}. None of the systems reach synchronization within the sdB lifetime and so the dynamical tide cannot explain any observed tidal synchronization. This approach assumes that all sdB stars have the same $E_2$, which has been demonstrated to be incorrect and will be addressed further in Sec. 4.3.

\subsection{Basic Convective Dissipation}
\begin{figure*}
	\includegraphics[width=\textwidth]{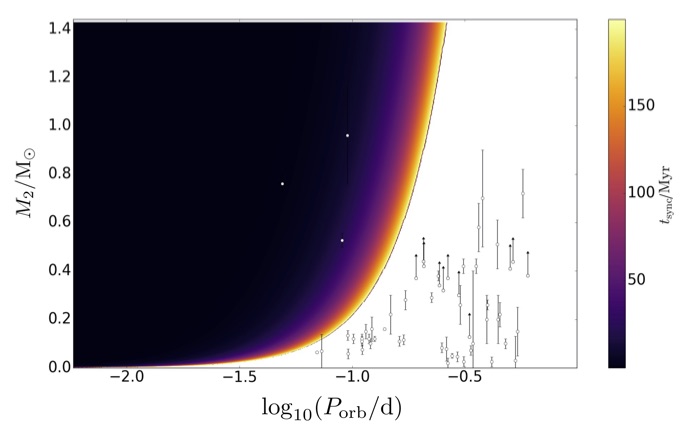}
    \caption{Plot of synchronization time-scales by convective dissipation for a single ZAEHB sdB model with $M_{\rm{sdb}}\,=\,0.47\, \rm{M_\odot}$, $M_{\rm{env}}\,=\,10^{-4} \,\rm{M_\odot}$. Companion masses less than the Chandrasekhar mass limit and orbital periods less than $0.5\,\rm{d}$ were calculated. Observed stars are open points with error bars. The synchronization time is shown in the colour bar. Only the closest sdBs with intermediate-mass white dwarf companions are expected to synchronize within the sdB lifetime.}
    \label{fig:contour}
\end{figure*}

We consider convective dissipation in the model of a zero-age extended horizontal-branch (ZAEHB) star with $M_{\rm{sdb}}\,=\,0.47\, \rm{M_\odot}$ and  $M_{\rm{env}}\,=\,10^{-4} \,\rm{M_\odot}$.
As a preliminary investigation into the significance of convective dissipation, the $P_{\rm{orb}}$ and $M_2$ parameter space for tidal synchronization within the EHB lifetime for a single sdB model has been computed and is shown in Fig. \ref{fig:contour}. This plot shows the results for convective dissipation assuming an initial rotation period of $100\,\rm{d}$ based on observations of rotation rates of red clump stars \citep{redclumprot}. As can be seen, synchronization is not achieved within the EHB lifetime by this mechanism except for the shortest period systems with relatively high-mass white dwarf companions. These systems rapidly reach a state of tidal synchronization.
The model was selected because it has the canonical mass of $0.47\,\rm{M_\odot}$ and one of the lower-mass envelopes of the grid. Owing to its influence on the overall stellar radius, the envelope mass is the main factor governing the fractional radial extent of the convective region. A model with a low envelope mass synchronizes more quickly than one with a higher envelope mass. Closer examination of the mixing length and velocity of this model shows that the convective turnover time is substantially longer than the orbital period. This results in the tidal forces being significantly damped because convective elements do not travel over the full mixing length during one orbital revolution \citep{chrissuggestion}. The implications of this damping are discussed the next section. 

\subsection{Synchronization Time-scales}
Convective dissipation approximates the convective viscosity as $\nu\,=\,wl$ where $w$ is the local velocity of MLT convective cells and $l$ is the size of these cells. The convective turnover time was found to be orders of magnitude longer than the orbital periods at which tides are most effective. This means the dissipation of the tides is damped. The damping factor is introduced to the equation for finding the convective time-scale of the tides and Eq. 9 is updated to
\begin{equation}
    \frac{1}{\tau_{\rm{visc}}}\,=\,\frac{1}{M_1R^2_1}\int_0^{M_{1}}wl\gamma(r)\Psi(r) dm.
\end{equation}
\cite{tidedampchris} define the damping factor $\Psi (r)$ as 
\begin{equation}
\Psi_1 (r)\,=\,\big|\frac{P_{\rm{orb}}}{2t_{\rm{turnover}}}\big|.
\end{equation}
\cite{zahn66} had previously introduced
\begin{equation}
\Psi_2 (r)\,=\,\big|\frac{P_{\rm{orb}}}{2t_{\rm{turnover}}}\big|^2.
\end{equation}
\cite{hydrodis} carried out 3D hydrodynamical simulations to investigate this damping and obtained  results  more in agreement with $\Psi_1(r)$. They suggest that these results are most applicable to tidal dissipation in gaseous planets owing to uncertainties in stellar convection feedback. We compare the two cases here. 

The mixing length in the core of the star tends to infinity when defined as $l\,=\alpha P/\rho g$ where $\alpha$ is the mixing length parameter, $P$ is the pressure, $\rho$ is the density and $g$ is the gravity. It is unphysical for $l$ to exceed the radius of the convective region so several different approximations were used to study the effect on the tidal synchronization times. The four mixing lengths examined are as follows: 
\begin{enumerate}
\item MLT1 is $l$ as predicted by traditional mixing length theory
\item MLT2 is  $l$ restricted to the distance to the edge of the convective region. This is the most necessary  constraint.
\item MLT3 has $l$ limited so that the convective turnover time is just less than the orbital period and so the tides are not damped. Assuming $l=r_{\rm{conv}}/20$ satisfies this.
\item MLT4 has $l=r_{\rm{conv}}/50$.

\end{enumerate}

\subsubsection*{\cdthirty}
\begin{figure*}
	\includegraphics[width=\textwidth]{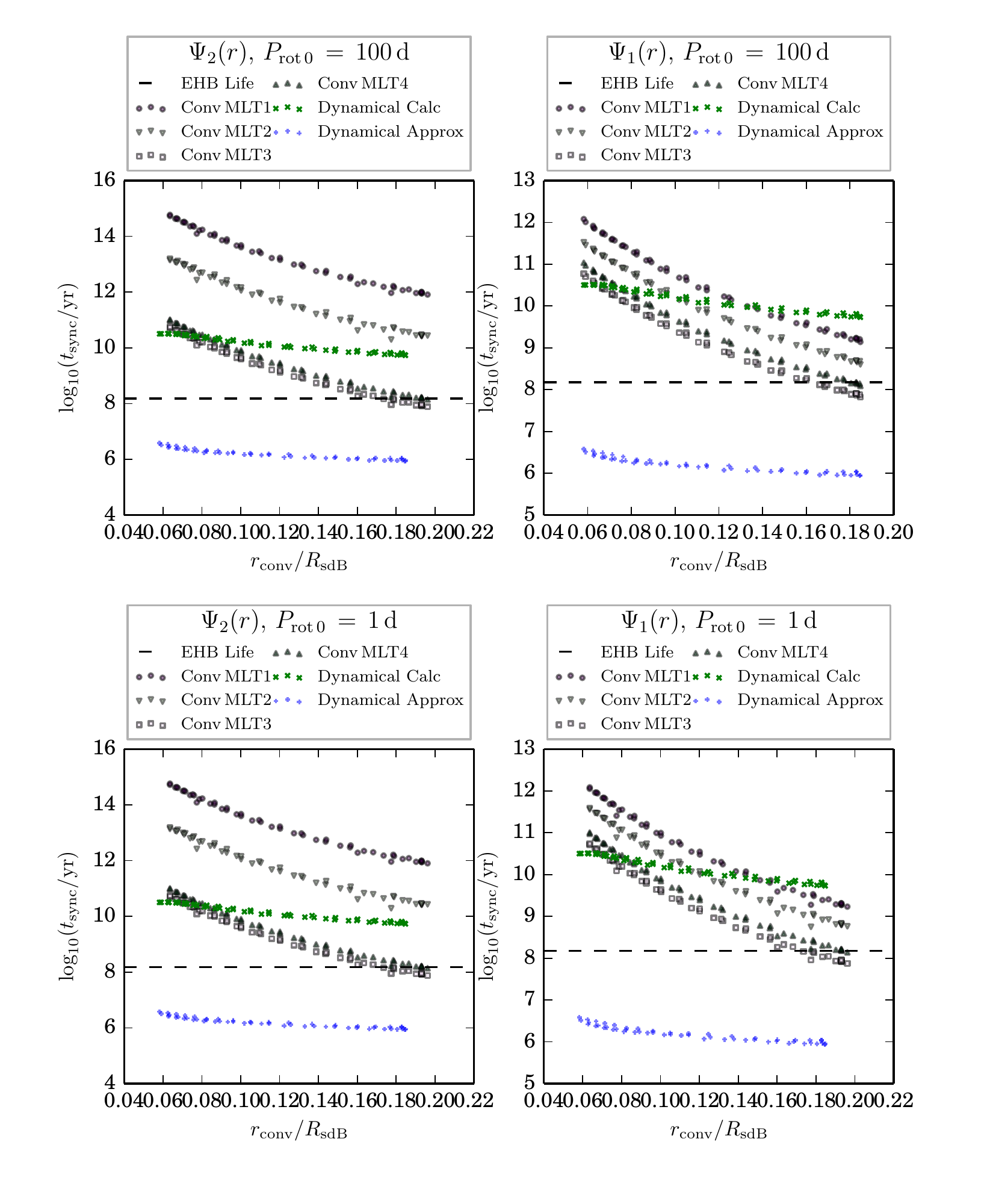}
    \caption{The synchronization time for \cdthirty\ as a function of the fractional size of the convective region for the ZAEHB models. The different mixing length (MLT) prescriptions are defined in Sec. 4.3. The initial rotation period ($P_{\rm{rot\,0}}$) and choice of damping factor ($\Psi_1(r)$ or $\Psi_2(r)$) are shown in the legend for each panel. MLT3 predicts the shortest time-scales whilst MLT1 predicts the longest. Dynamical calc refers to Zahn's calculations with $E_2$ calculated as discussed in the text. Dynamical approx refers to Zahn's calculations with $E_2\,=(r_{\rm{conv}}/R_{\rm{sdB}})^8$.}
    \label{figcd-30tsync}
\end{figure*}

The effects of initial rotation rate and different damping factors are considered for the full set of ZAEHB models and applied to the \cdthirty\ system in Fig. \ref{figcd-30tsync}. Properties of $\cdthirty$ can be found in Table 1. The majority of models predict synchronization time-scales longer than the typical EHB lifetime. In the most efficient cases, with a modified MLT, synchronization via convective dissipation, is predicted within the EHB lifetime for some models. Even the models which predict synchronization within the EHB lifetime do so on times comparable to this evolutionary stage meaning assumptions of tidal synchronization should be made with extreme care. An initial rotation rate of 1$\,\rm{d}$ only has a very small effect on the synchronization time-scales.

The equilibrium tidal dissipation time-scales are generally longer than dynamical tide dissipations unless MLT3 or MLT4 are used as can be seen in Fig.~\ref{figcd-30tsync}. Calculation by \cite{geier2010}'s method predicts synchronization well within the EHB lifetime of $150\,\rm{Myr}$. However detailed calculation of $E_2$ does not predict this system to be synchronized. The results are slightly different to those predicted in section 4.1 because $E_2$ is calculated individually for each model.

Without taking damping of the tidal dissipation into consideration, shorter mixing lengths predict longer synchronization times because the viscosity in the convective region is smaller. However, the fastest synchronization predictions are for MLT3 because the convective turnover time for this scheme is just below the threshold for the damping to be applied. If the mixing length is longer than this the dissipation of the tides are damped and if the mixing length is shorter the viscosity decreases. When $\Psi_1(r)$ is used the dependence on the mixing length is decreased for MLT1 and MLT2. MLT3 and MLT4 have convective turnover times shorter than the orbital period and so are not affected by the damping. MLT4 predicts slightly longer synchronization times than MLT3 because it has a lower viscosity. 

\begin{figure}
	\includegraphics[width = \columnwidth]{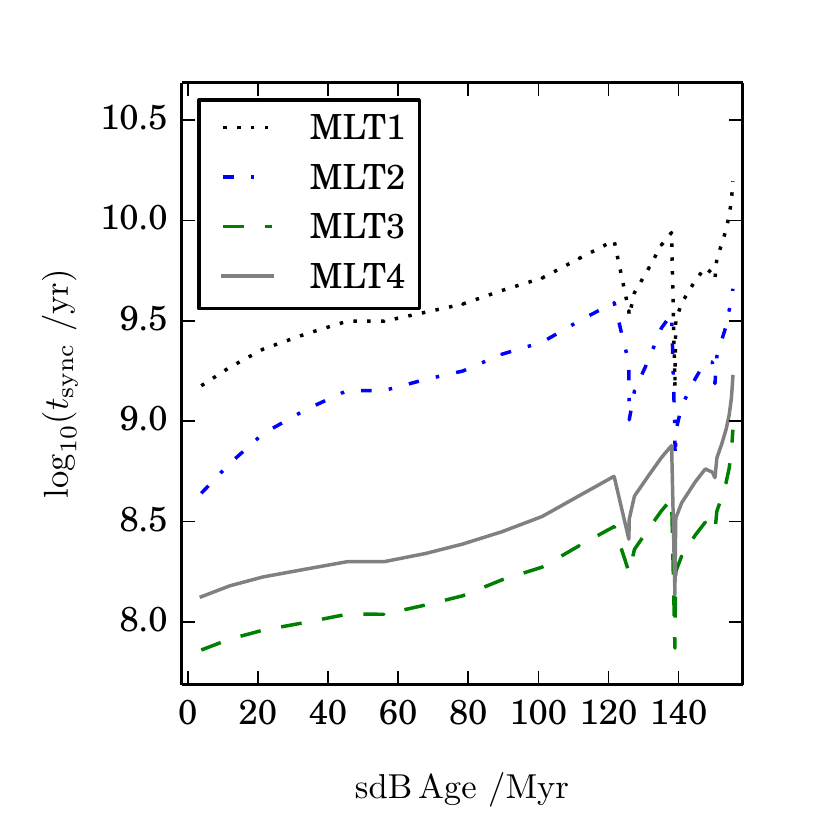}
    \caption{The synchronization time as a function of sdB age for a 0.47$\,\rm{M_\odot}$ sdB star with a $10^{-4}\,\rm{M_\odot}$ envelope. An initial rotation period of 1$\,\rm{d}$ and a damping factor $\Psi(r)_1$ were used.}
    \label{tvt}
\end{figure}

The synchronization time as a function of sdB age for a 0.47$\,\rm{M_\odot}$ sdB star with a $10^{-4}\,\rm{M_\odot}$ envelope can be seen in Fig.~\ref{tvt}. The ZAEHB models predict the shortest synchronization time-scales. For these calculations a damping factor of $\Psi(r)_1$ and an initial rotation rate of 1$\, \rm{d}$ were used.

\subsection{Change in Rotational Period Over sdB Lifetime}
At this stage, it is apparent that sdB stars do not synchronize in the EHB lifetime. Despite this, the tides may still cause the stars to be spun up to some degree. The change in the angular velocity $\Omega (t)$ of the sdB star as it evolves can be calculated by integrating Eq. 12 with $k_r$, $R$ and $\tau_{\rm{tide}}$ are all functions of time. 

\begin{figure*}
	\includegraphics[width=0.90\textwidth]{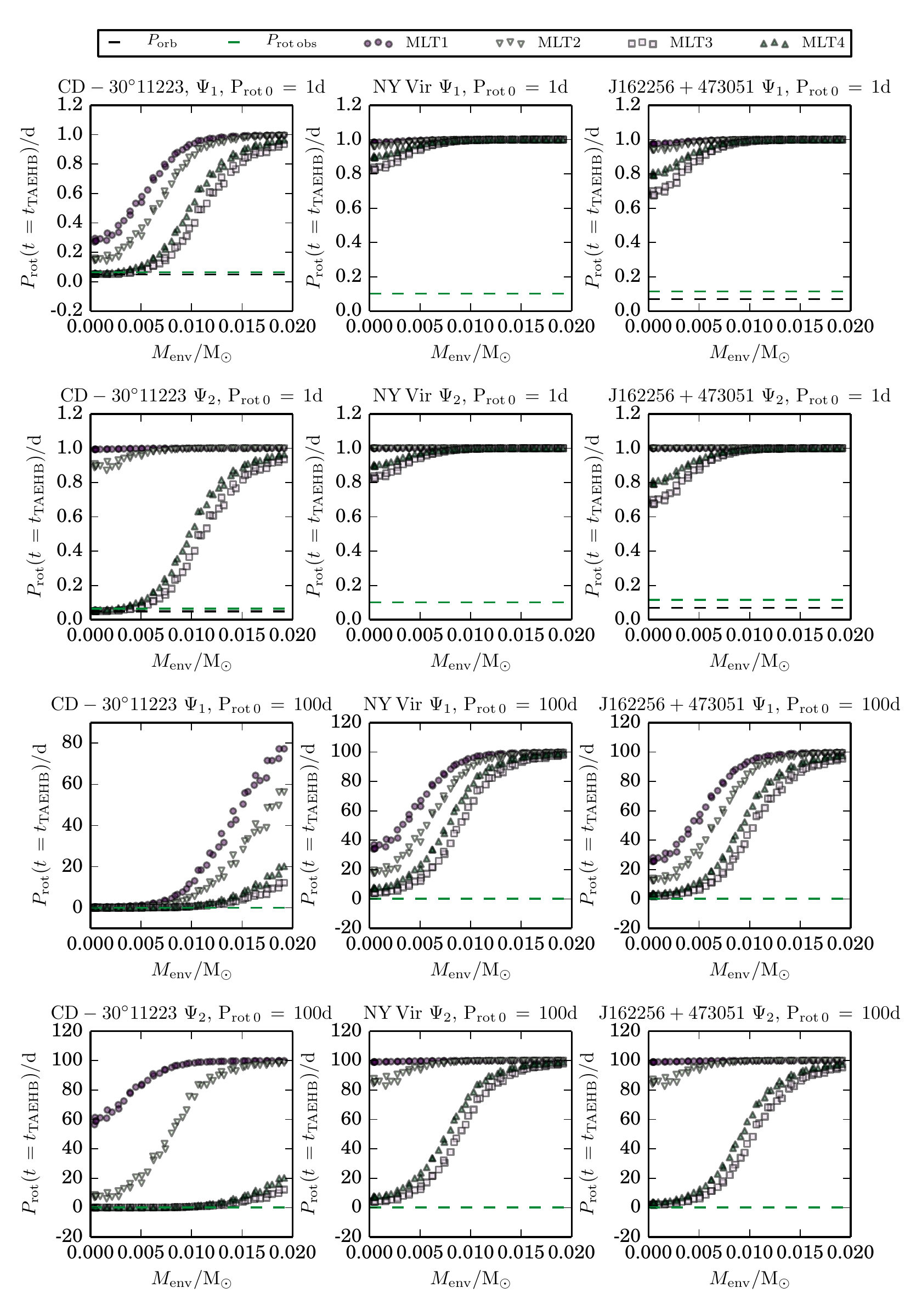}
    \caption{The rotational period of the sdB evolutionary sequences at the TAEHB is shown as a function of the envelope mass of the model. The MLT and damping prescriptions are described in Sec. 4.3. Whilst the envelope mass does not directly influence the radius of the convection zone it is the main factor governing the stellar radius $R_{\rm{sdB}}$. Tidal interactions are strongly dependent on $r_{\rm{conv}}/R_{\rm{sdB}}$. The data points shown here are for evolutionary sequences with an $r_{\rm{conv}}/R_{\rm{sdB}}$ which varies. }
    \label{omegat}
\end{figure*}

These calculations were applied to the systems \cdthirty\,, \jsixteen\ and NY\,Vir to find the rotational period at the TAEHB. \jsixteen\ is the shortest period sdB binary not observed to be tidally synchronized. It has $P_{\rm{orb}}\,=\,0.069\,\rm{d}$ and sub-stellar companion mass $M_2\,=\,0.064\,\rm{M_\odot}$.  Neither convective dissipation nor radiative dissipation predict the synchronization of this system. NY\,Vir is the only sdB with asteroseimsological evidence suggesting that it is rotating synchronously. The rotational period at the TAEHB was calculated using $\Psi_1(r)$ and $\Psi_2(r)$ and multiple mixing length schemes and can be seen in Fig. \ref{omegat}.

In contradiction to the observations, \jsixteen\ is predicted to be spun up more than NY\,Vir by the time it reaches the TAEHB. This is due to the fact that the orbital period of \jsixteen\ is substantially shorter than that of NY\,Vir. 
Using traditional mixing length theory or restricting the mixing length to the Schwarzschild radius, we find the stars not to be spun up at all. If the mixing length is limited so that the convective turnover time is faster than the tides and the tidal forces are no longer dissipated, all three systems considered are spun up to some degree. If $\Psi_1(r)$ is used the mixing length dependence becomes less strong. MLT3 and MLT4 are independent of $\Psi(r)$ because they have sufficiently fast convective turnover times.

\subsection{Convective Cores and Associated Uncertainties}
The models presented above use standard convection
theories widely implemented in stellar evolution codes. However, the extent
of the convective core measured in some asteroseismic studies, $\log (1 - m_{\rm{conv}}/M_{\rm{sdB}})= -0.30$ \citep{vangrootel2010b, charpinet2011}, is somewhat larger than that seen in our models. Additional evidence from white dwarf asteroseismology \citep{naturecores} suggests even more of the core of the post horizontal-branch star has been homogenized, presumably by additional convective processes. Evidence suggests that red clump stars also have larger convective regions than predicted by standard stellar models \citep{redclumpcoresize}. A newly adopted maximal overshooting scheme must be used to reproduce the period spacings of the g-dominated mixed modes observed in these stars. However the physical validity of such a scheme is still in question. 

Recent theoretical investigations show that extreme care must be taken when determining edges of convective regions in stellar evolution codes \citep{gabrielconvboundary,mesaconvboundary}. Both studies find that the exact method used to find the convective boundary has consequences for the subsequent evolution of a model.

In summary, the physics of helium burning cores is still not well established. In the context of tidal interactions, a larger convective core mass implies a larger fractional convective core radius and hence a shorter tidal synchronization time. However, in the absence of a self-consistent framework in which  to compute extreme-horizontal branch models  with larger convective cores, it is not possible to compute the effect directly. A parametric investigation would make a worthwhile study.


\section{Conclusions}
The goal of this study was to find synchronization time-scales for short period sdB binary systems. A grid of sdB models was created with the STARS code for a variety of progenitor masses, envelope masses and treatments of convection. Previous studies have predominantly used Zahn's theory of dynamical tides with a scaling from main-sequence models to find the synchronization times. Recalculating the tidal coefficient $E_2$ for the grid of sdBs shows scaling from main-sequence models overpredicts $E_2$ by a factor of at least 3000. The synchronization time-scales should be several orders of magnitude longer. As a result, estimates of Zahn's dynamical tide synchronization time-scales are longer than EHB lifetimes, even for the extreme case of \cdthirty. 

The sdB stars have convective cores which provide a mechanism for tidal dissipation. By solving Clairaut's equation the tidal synchronization times owing to turbulent convection have been calculated. Initial calculations of the convective tides predicted that the three sdB systems with the most massive WD companions should be synchronized. Closer examination revealed that the orbital period is typically shorter than the convective turnover time. This causes the convective dissipation of the tides to be damped and become substantially less efficient. The damping coefficient depends on the turnover time for viscous elements within the star and is calculated with mixing length theory. The damping factor causes estimates of synchronization time-scales to increase by several orders of magnitude so that no sdB binary systems are conclusively predicted to be synchronized. 

Traditional mixing length theory predicts a singularity at the stellar centre.  The effects on tidal synchronization time-scales when the mixing length was altered to remove this singularity were examined. Reducing the mixing length to avoid the central singularity generally increases the synchronization time because the estimated viscosity decreases. The optimal case for tidal dissipation is to reduce the mixing such that the convective turnover time is slightly shorter than the orbital period so that the tidal dissipation is not damped. Even in this case synchronization is not achieved because the viscosity is substantially reduced and the tidal interactions are less efficient. 

The rotational periods of sdB stars at the TAEHB were calculated to investigate the impact of the tides. The models with the optimally chosen mixing length and with envelope masses less than $0.01\,\rm{M_\odot}$ are most substantially affected by the tides. The convective region accounts for a larger fractional volume in the sdBs with the lowest mass envelopes so tides are more effectively dissipated.

With the theoretical framework presented, tidal synchronization times for EHB stars are long, but not excessively so, compared with nuclear lifetimes. With evidence from asteroseismology that convective core sizes may be larger than those predicted by classical convection theory, and with the possibility that the tides could induce differential rotation with the EHB star, these avenues of exploration still open.

\section*{Acknowledgements}
Research at the Armagh Observatory and Planetarium is supported by a grant-in-aid from the Northern Ireland Department for Communities. HPP acknowledges support from  the UK Science and Technology Facilities Council (STFC) Grant No. ST/M502268/1. CSJ acknowledges support from STFC Grant No. ST/M000834/1. CAT thanks Churchill College for his fellowship.



\bibliographystyle{mnras}
\bibliography{references} 







\bsp	
\label{lastpage}
\end{document}